\documentclass{article}

 \usepackage[position, final]{neurips_2026}

\usepackage[utf8]{inputenc} 
\usepackage[T1]{fontenc}    
\usepackage{hyperref}       
\usepackage{xurl}            
\usepackage{booktabs}       
\usepackage{amsfonts}       
\usepackage{nicefrac}       
\usepackage{microtype}      
\usepackage{xcolor}         

\usepackage{colortbl, booktabs}
\definecolor{lightgray}{gray}{0.9}
\definecolor{lightgreen}{RGB}{200, 255, 200}
\definecolor{lightred}{RGB}{255, 200, 200}
\definecolor{lightblue}{RGB}{230, 240, 255}
\definecolor{lightyellow}{RGB}{255, 250, 220}
\usepackage{amsmath}
\usepackage{amssymb}
\usepackage{mathtools}
\usepackage{amsthm}

\usepackage{tikz}
\usetikzlibrary{shapes, arrows.meta, positioning, shadows}
\usepackage{graphicx}
\usepackage{multirow} 
\usepackage{adjustbox} 
\usepackage{rotating}
\usepackage{array}
\newcolumntype{C}[1]{>{\centering\arraybackslash}p{#1}}

\title{Watermarking Without Standards Is Not AI Governance}

\author{%
  Alexander Nemecek, Yuzhou Jiang, Erman Ayday \\
  Case Western Reserve University \\
  \texttt{\{ajn98, yxj466, exa208\}@case.edu} \\
}

\begin{document}

\maketitle

\begin{abstract}
Watermarking has emerged as a leading technical proposal for attributing generative AI content and is increasingly cited in global governance frameworks. This position paper argues that current implementations risk serving as symbolic compliance rather than delivering effective oversight. We identify a growing gap between regulatory expectations and the technical limitations of existing watermarking schemes. Through analysis of policy proposals and industry practices, we show how incentive structures disincentivize robust, auditable deployments. To realign watermarking with governance goals, we propose a three-layer framework encompassing technical standards, audit infrastructure, and enforcement mechanisms. Without enforceable requirements and independent verification, watermarking will remain inadequate for accountability and ultimately undermine broader efforts in AI safety and regulation.
\end{abstract}

\section{Introduction}\label{intro}
\begin{quote}
    \textit{``A law without teeth is just a suggestion.''}
\end{quote}
This adage captures a growing concern in AI governance, where policies are advancing more quickly than the technical tools available to enforce them. A prominent example of this mismatch is \textit{watermarking}, a family of techniques designed to embed identifiable signatures into AI-generated content. Watermarking is widely referenced in AI governance frameworks, yet current implementations often lack the robustness, transparency, and verifiability required to fulfill this role effectively.

Recent advances in generative AI have significantly increased the scale and realism of synthetic content, including text, images, and audio~\cite{spennemann2025delving, fisher2024moderating}. As such content is deployed in sensitive areas like education, healthcare, and finance, policymakers have identified attribution and provenance as urgent challenges. Misattributed content contributes to the spread of misinformation and to technical failures such as feedback loops in model training, including contamination and eventual model collapse~\cite{Shumailov2024}.

In response, watermarking has gained traction in both technical research~\cite{kirchenbauer2023watermark, fernandez2023stable, chen2023wavmark} and policymaking. The United States Executive Order 14110, for example, mandates ``state-of-the-art'' provenance tools and cites watermarking explicitly~\cite{ExecOrder14110}. The European Union's AI Act requires machine-readable content markings~\cite{EUAIAct2024}, with similar provisions being proposed in jurisdictions worldwide~\cite{zhao2024sok}. Although watermarking is just one of several provenance strategies, it occupies an important role in current governance discourse.

This reliance rests on a flawed foundation. While watermarking techniques vary across modalities, text and audio schemes remain brittle (Table~\ref{tab:watermark-eval}), and most remain difficult to audit and proprietary. Policymakers often assume these methods can be standardized and verified. In practice, industry deployments obscure technical details while asserting compliance, turning watermarking into a box-checking exercise rather than a meaningful tool. Lacking common standards, evaluation infrastructure, or defined threat models, current implementations are unlikely to fulfill their intended governance role.

To address this growing disconnect, \textbf{we advance two central positions:}
\begin{itemize}
    \item \textbf{Watermarking schemes must be designed with verifiability and auditability as primary technical requirements, rather than implemented as proprietary black boxes.}
    \item \textbf{Policymakers must require technical standards and independent testing to ensure that watermarking fulfills governance goals in practice.}
\end{itemize}
\noindent To support these positions, we make three contributions. First, we analyze emerging watermarking mandates and demonstrate they assume levels of technical feasibility that current systems do not meet. Second, we examine the incentive structures driving industry watermarking deployments, showing how market dynamics often disincentivize robust, auditable implementations. Third, we propose a three-layer framework, spanning technical design, audit infrastructure, and enforcement which realigns the design of watermarking systems with their intended governance functions.

While prior work has analyzed watermarking's technical constraints, our contribution reframes these issues through a governance lens and provides a structured, actionable framework for aligning technical design with regulatory intent. Without intervention at the intersection of policy and design, watermarking is unlikely to deliver meaningful accountability. Instead, it risks becoming a symbolic substitute for the more demanding components of effective AI regulation.

\section{Governance Aspirations vs. Technical Realities}\label{govVtech}
As watermarking becomes a feature of AI governance proposals, many mandates rely on technical assumptions that current systems cannot fulfill. We identify three such assumptions and contrast each policy aspiration with technical realities, highlighting key gaps that must be addressed for watermarking to serve as an effective governance tool.

\subsection{Assumption I: Watermarking Robustness Can Be Required Without Being Defined}
Many governance proposals call for watermarking that is robust against tampering, yet leave undefined what level of robustness satisfies this requirement. Whether applied to text, images, or audio, watermarking is expected to remain detectable throughout the content lifecycle, but no threshold specifies how much degradation is acceptable.

\begin{quote}
\textit{``The disclosure is permanent or extraordinarily difficult to remove, to the extent it is technically feasible.''} --- California Senate Bill No. 942~\cite{SB942}
\end{quote}

\begin{quote}
\textit{``Such techniques and methods should be sufficiently reliable, interoperable, effective and robust as far as this is
technically feasible, taking into account available techniques or a combination of such techniques, such as
watermarks,\dots as may be appropriate.''} --- European Union AI Act~\cite{EUAIAct2024}
\end{quote}

These documents reflect the shared policy expectation of robustness but delegate its meaning to phrases such as `technically feasible.' The delegation is standard as quantitative thresholds belong in implementing standards and codes of practice that laws adopt by reference. For example, the Americans with Disabilities Act names no web standard, yet its implementing rules require conformance to the Web Content Accessibility Guidelines~\cite{ADA1990, DOJWebRule2024, WCAG21} and the Clean Air Act sets no pollutant levels, but directs the Environmental Protection Agency to set them~\cite{CAA7409, NAAQS40CFR50}. For watermarking, no layer of this pipeline yet defines robustness quantitatively. The EU's own Code of Practice on transparency does not name a marking technique because under the current state of the art no single technique meets all four of the properties Article 50 asks for. This leaves providers to test with their own methods until common benchmarks exist~\cite{ECCoP2026, garina2026}. California has moved the operative date of its Transparency Act to line up with the EU~\cite{AB853}, so both jurisdictions are now waiting on the same missing piece. Without such thresholds or shared metrics for evaluating robustness under adversarial conditions, meaningful compliance remains unverifiable and enforcement difficult.

In practice, watermarking schemes often fall short of this robustness ideal. Google's SynthID~\cite{synthid2023}, for example, embeds imperceptible watermarks in AI-generated text. While resilient to certain distortions, its detection can be evaded by simple edits such as character perturbations or short-form text paraphrasing~\cite{dathathri2024scalable, nemecek2024topic}. Across all modalities, there is currently no standardized evaluation protocol for watermark robustness, and few systems offer guarantees under adversarial or worst-case conditions.

The absence of shared benchmarks or test suites for robustness creates two problems: first, it makes meaningful regulatory compliance unverifiable; second, it enables firms to implement weak watermarking schemes while claiming alignment with policy language. 

To concretely illustrate this gap in industry, we conducted minimal robustness evaluations across four modalities using industry watermarking systems: Google's SynthID-Text~\cite{dathathri2024scalable} for text, Meta's AudioSeal~\cite{roman2024proactive} for audio, PixelSeal~\cite{soucek2025pixelseal} for images, and VideoSeal~\cite{fernandez2024videoseal} for video. SynthID-Text is deployed in Gemini, but direct access to detection scores is unavailable; we therefore evaluate on \texttt{Gemma-2-9b-it}~\cite{team2024gemma}, Google's open-source model family. Similarly, AudioSeal has been deployed in Meta's public-facing AI products as a post-hoc watermarking system applied to generated audio~\cite{fernandez2025audioseal}, but we evaluate on \texttt{AudioGen}~\cite{kreuk2022audiogen} as an accessible proxy. No deployed image or video scheme publicly exposes detection scores, so for those modalities we evaluate PixelSeal and VideoSeal, Meta's open post-hoc schemes, on outputs of \texttt{SDXL}~\cite{podell2024sdxl} and \texttt{CogVideoX-2b}~\cite{yang2024cogvideox}.

For each modality, we generated five outputs and evaluated three variants. Text outputs were tested in original, paraphrased, and summarized forms, with transformations applied using GPT-4o~\cite{hurst2024gpt}. Audio and video outputs were tested in original form, with modified playback speed, and with additive noise, and image outputs in original form, after JPEG compression, and with additive noise. These transformations reflect common user modifications across modalities. Detection thresholds for SynthID-Text and AudioSeal were obtained from published documentation~\cite{dathathri2024scalable, roman2024proactive}; PixelSeal and VideoSeal publish none, so we use the smallest bit accuracy with $p \leq .01$ against the unwatermarked null for a 256-bit payload. As shown in Table~\ref{tab:watermark-eval}, the results split by modality: for text and audio, even benign edits degraded detection below threshold, with 9 of 10 original outputs detected but only 1 of 20 transformed outputs, whereas for images and video all 10 original and 19 of 20 transformed outputs remained detectable. Full evaluation details, prompts, and outputs are provided in Appendix~\ref{appendix:eval}. Industry made the same point in the hearings on AB 853, arguing that provenance data ``may degrade or be lost for benign technical reasons (e.g., compression or transcoding)''~\cite{AB853analysis}. What Table~\ref{tab:watermark-eval} shows for text and audio is now an argument in the drafting of the law. The split shows that watermarking is not too fragile to govern, but no shared bar yet separates the schemes that survive routine edits from those that do not (Section~\ref{three-layer-framework}).

\begin{table}[h]
\centering
\small
\caption{Watermark detection under benign edits for Google's SynthID-Text (threshold $\geq$0.52), Meta's AudioSeal ($\geq$0.5), PixelSeal ($\geq$0.578), and VideoSeal ($\geq$0.578). Green marks values at or above threshold (watermark detected). PixelSeal and VideoSeal scores are bit accuracy, with the binomial threshold derived in Appendix~\ref{appendix:eval}.}
\label{tab:watermark-eval}
\begin{tabular}{@{}lccc@{}}
\toprule
\rowcolor{lightgray}
\textbf{SynthID-Text} & Original & Paraphrase & Summarize \\
\midrule
$\text{Sample}_1$ & \cellcolor{lightgreen}.520 & \cellcolor{lightred}.489 & \cellcolor{lightred}.515 \\
$\text{Sample}_2$ & \cellcolor{lightgreen}.527 & \cellcolor{lightred}.493 & \cellcolor{lightred}.497 \\
$\text{Sample}_3$ & \cellcolor{lightgreen}.521 & \cellcolor{lightred}.518 & \cellcolor{lightred}.513 \\
$\text{Sample}_4$ & \cellcolor{lightred}.505 & \cellcolor{lightred}.501 & \cellcolor{lightred}.511 \\
$\text{Sample}_5$ & \cellcolor{lightgreen}.526 & \cellcolor{lightred}.509 & \cellcolor{lightred}.487 \\
\midrule
\rowcolor{lightgray}
\textbf{AudioSeal} & Original & Speed & Noise \\
\midrule
$\text{Sample}_1$ & \cellcolor{lightgreen}1.00 & \cellcolor{lightred}.000 & \cellcolor{lightgreen}.964 \\
$\text{Sample}_2$ & \cellcolor{lightgreen}1.00 & \cellcolor{lightred}.000 & \cellcolor{lightred}.022 \\
$\text{Sample}_3$ & \cellcolor{lightgreen}1.00 & \cellcolor{lightred}.012 & \cellcolor{lightred}.001 \\
$\text{Sample}_4$ & \cellcolor{lightgreen}1.00 & \cellcolor{lightred}.003 & \cellcolor{lightred}.418 \\
$\text{Sample}_5$ & \cellcolor{lightgreen}1.00 & \cellcolor{lightred}.000 & \cellcolor{lightred}.087 \\
\bottomrule
\end{tabular}
\hspace{1.5em}
\begin{tabular}{@{}lccc@{}}
\toprule
\rowcolor{lightgray}
\textbf{PixelSeal} & Original & JPEG & Noise \\
\midrule
$\text{Sample}_1$ & \cellcolor{lightgreen}1.00 & \cellcolor{lightgreen}1.00 & \cellcolor{lightgreen}.734 \\
$\text{Sample}_2$ & \cellcolor{lightgreen}1.00 & \cellcolor{lightgreen}.969 & \cellcolor{lightgreen}.613 \\
$\text{Sample}_3$ & \cellcolor{lightgreen}.996 & \cellcolor{lightgreen}.965 & \cellcolor{lightgreen}.652 \\
$\text{Sample}_4$ & \cellcolor{lightgreen}1.00 & \cellcolor{lightgreen}.988 & \cellcolor{lightgreen}.738 \\
$\text{Sample}_5$ & \cellcolor{lightgreen}1.00 & \cellcolor{lightgreen}.996 & \cellcolor{lightgreen}.684 \\
\midrule
\rowcolor{lightgray}
\textbf{VideoSeal} & Original & Speed & Noise \\
\midrule
$\text{Sample}_1$ & \cellcolor{lightgreen}1.00 & \cellcolor{lightgreen}1.00 & \cellcolor{lightgreen}.664 \\
$\text{Sample}_2$ & \cellcolor{lightgreen}.988 & \cellcolor{lightgreen}.984 & \cellcolor{lightred}.504 \\
$\text{Sample}_3$ & \cellcolor{lightgreen}.996 & \cellcolor{lightgreen}1.00 & \cellcolor{lightgreen}.727 \\
$\text{Sample}_4$ & \cellcolor{lightgreen}1.00 & \cellcolor{lightgreen}.996 & \cellcolor{lightgreen}.910 \\
$\text{Sample}_5$ & \cellcolor{lightgreen}1.00 & \cellcolor{lightgreen}1.00 & \cellcolor{lightgreen}.703 \\
\bottomrule
\end{tabular}
\end{table}

\subsection{Assumption II: Watermarking Enables Independent and Reliable Detection}\label{ass2}
A second assumption in many governance proposals is that watermarking will enable independent third parties (e.g., regulators, researchers, platform operators) to reliably detect AI-generated content. This assumption seeks to enforce provenance or hold actors accountable for synthetic content.

\begin{quote}
\textit{``[Develop] Methods for the verification of statements of digital content provenance to ensure authenticity such as watermarking or classifiers, which are trained models that distinguish artificial intelligence-generated content.''} --- United States Congress S. 3312~\cite{ai_accountability_act_2023}
\end{quote}

\begin{quote}
\textit{``Deep synthesis service providers shall fulfill entity responsibility for information security, establish and complete management structures for algorithmic mechanism and logic examination and verification\dots with secure and controllable technical safeguard mechanisms suited to the development of new technologies and new applications.''} --- China's Deep Synthesis Provisions~\cite{CAC}
\end{quote}

These statements imply that watermarking should support reliable, externally verifiable detection. Most watermarking systems today do not support this expectation due either to limited transparency or configuration dependence.

For instance, OpenAI withheld its text watermark~\cite{wsj2024watermark}; the watermark it deployed in 2026 is EU-only or opt-in, with detector access by application and promised public implementation~\cite{openai2026textgrain}. Google's SynthID-Text~\cite{dathathri2024scalable} goes further, with watermarking and detection in public tools, but detection requires the generation configuration, which for deployed models is available only by application.

While the technical foundations for third-party detection are emerging, the practical reality remains constrained. Most watermarking approaches do not currently support universal, auditable detection, and governance frameworks often overlook the infrastructure and standardization needed to make detection viable at scale.

\subsection{Assumption III: Industry Will Voluntarily Align with Governance Goals}\label{ass3}
A final assumption in governance frameworks, particularly in the United States and European Union, is that generative AI providers will voluntarily adopt watermarking practices aligned with policy goals.

\begin{quote}
\textit{``\dots establish a strong international framework for managing the risks and harnessing the benefits of AI, including by encouraging international allies and partners to support voluntary commitments similar to those that United States companies have made in pursuit of these objectives\dots to develop common regulatory and other accountability principles for foreign nations, including to manage the risk that AI systems pose.''} --- United States Executive Order 14110~\cite{ExecOrder14110}
\end{quote}

\begin{quote}
\textit{``Even though adherence to the code is voluntary, the transparency requirements under article 50 of the AI Act are legal obligations.''} --- European Commission, Code of Practice on Transparency of AI-Generated Content~\cite{ECCoP2026}
\end{quote}

This reflects a broader strategy of governance-by-consensus, based on reputational pressure and shared norms. Symbolic compliance under such strategies is a documented pattern across regulated industries~\cite{edelman1992legal}; watermarking is where it becomes testable, since the mandates name a measurable artifact. The United States Biden-Harris Administration secured voluntary commitments from leading AI companies explicitly referencing watermarking to lead safe and transparent AI~\cite{whitehouse2023aicommitments, whitehouse2023aicommitments2}.

However, implementation remains fragmented, with many deployments being firm-specific and lacking interoperability. Few are publicly auditable or developed through shared infrastructure. In the absence of enforceable standards or independent oversight, these commitments risk enabling symbolic compliance, signaling safety while delivering little practical governance capability.

Voluntary alignment also depends on political continuity. In 2025, the United States Trump-Vance Administration rescinded Executive Order 14110, directing agencies to ``suspend, revise, or rescind'' related initiatives~\cite{whitehouse2025aiorder}. The administration's stated focus on technological competitiveness leaves the governance status of watermarking ambiguous, casting uncertainty over the future of watermarking as a governance tool. The same administration's AI Action Plan still asks NIST to turn its Guardians of Forensic Evidence deepfake evaluation program into a formal guideline and a voluntary benchmark~\cite{AIActionPlan2025}. By contrast, China's Deep Synthesis Provisions~\cite{CAC} offer a more directive model, mandating compliance and audits, but even in such regimes, enforcement is uneven and robustness is not guaranteed.

While voluntary commitments may promote cooperation, they offer a fragile foundation for governance. When watermarking imposes costs or strategic risks, firms are unlikely to sustain alignment in the absence of legal or institutional compulsion.

\section{Why the Gap Persists: Industry Incentives}\label{incentives}
Despite growing regulatory attention, industry watermarking implementations remain fragmented. This is not simply a technical lag but reflects a deeper misalignment between governance objectives and industry incentives. Without enforceable standards, firms have strong incentives to pursue symbolic or minimal compliance~\cite{aaronson2023watermarking, edelman1992legal}.

First, watermarking offers reputational value. Companies can signal alignment with governance goals by announcing watermarking initiatives, often without disclosing technical details or enabling verification. This satisfies public demands at low cost, while sidestepping the risks of committing to truly robust solutions.

Second, robust watermarking entails economic and strategic risks. If detection tools are open source, they can be exploited by modifying outputs to evade detection or crafting content that falsely appears watermarked. Conversely, restricted-access to detection capabilities would require companies to develop secure infrastructure to manage watermarking keys, detection thresholds, and access control via cryptographic or trust-based frameworks. These trade-offs make firms hesitant to invest in public, auditable watermarking systems that support third-party verification.

Third, companies face disincentives to move unilaterally. Without coordination, early adopters risk losing users if watermarking is perceived as restrictive, such as for moderation or IP enforcement, especially if competitors offer unmarked alternatives. In a rapidly evolving market, the incentive to retain users outweighs alignment with long-term governance objectives.

Finally, regulatory uncertainty undermines long-term planning, further amplifying industry hesitation. As seen with the revocation of Executive Order 14110~\cite{whitehouse2025aiorder}, shifting political priorities can quickly alter the expected policy landscape. In such an environment, companies are hesitant to invest in watermarking infrastructure that may soon be obsolete.

These forces encourage minimal, firm-specific, and performative watermarking. The incentives are documented and remedies have been proposed, from private watermarks concentrating authorship determination in firms~\cite{raz2024authorship} to third-party audit ecosystems~\cite{raji2022outsider} and detection as a condition of release~\cite{knott2023generative}; what is new here is their synthesis into one enforceable chain, incentives to standards to audits to penalties, with a scorecard and audit infrastructure that make it testable.

\section{A Three-Layer Approach to Enforceable Watermarking}\label{three-layer-framework}
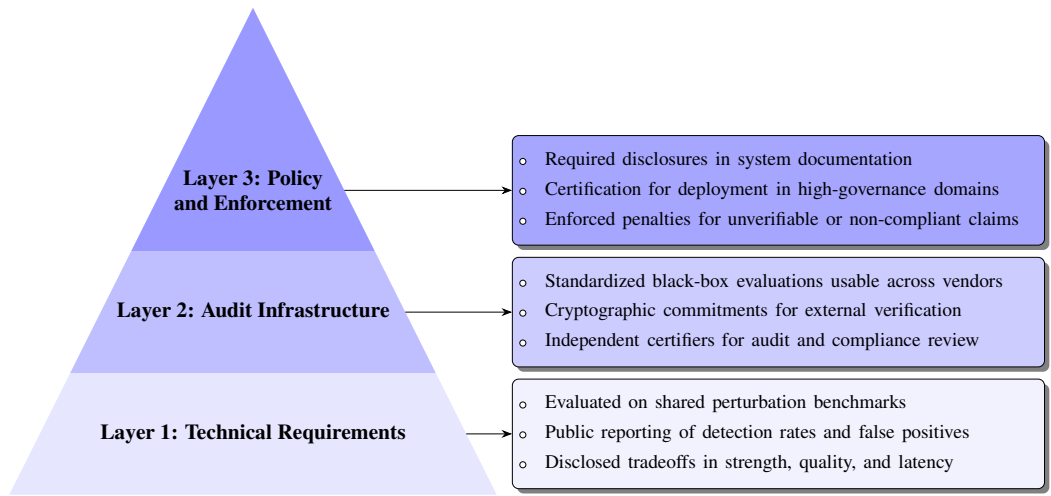
\begin{figure*}[h]
\centering
\scalebox{0.78}{
\begin{tikzpicture}[scale=1, every node/.style={align=left}, node distance=1cm and 4cm]

\fill[blue!10] (-4,0) -- (4,0) -- (3,2) -- (-3,2) -- cycle;
\node at (0,1.0) {\textbf{Layer 1: Technical Requirements}};

\fill[blue!25] (-3,2) -- (3,2) -- (2,4) -- (-2,4) -- cycle;
\node at (0,3) {\textbf{Layer 2: Audit Infrastructure}};

\fill[blue!40] (-2,4) -- (2,4) -- (0,8) -- cycle;
\node at (0,5) {\shortstack{\textbf{Layer 3: Policy}\\\textbf{and Enforcement}}};

\draw[-{Stealth}] (-3.75,0.5) to[bend left=12]
  node[midway, sloped, above, font=\scriptsize, inner sep=2pt] {evidence} (-1.85,4.3);
\draw[{Stealth}-, dashed] (-3.95,0.1) to[bend left=42]
  node[midway, sloped, above, font=\scriptsize, inner sep=2pt] {revision} (-1.45,5.1);

\node[draw, fill=blue!5, anchor=west, align=left, text width=8.5cm, minimum height=1.8cm, rounded corners=3pt, drop shadow] (techbox) at (4.25,1.0)
{{\footnotesize
\tikz\draw[fill=white] (0,0) circle (0.05cm); \hspace{0.5ex} Evaluated on shared perturbation benchmarks\\[1ex]
\tikz\draw[fill=white] (0,0) circle (0.05cm); \hspace{0.5ex} Public reporting of detection rates and false positives\\[0.75ex]
\tikz\draw[fill=white] (0,0) circle (0.05cm); \hspace{0.5ex} Disclosed tradeoffs in strength, quality, and latency 
}};
\draw[-{Stealth}] (3.5,1.0) -- (techbox.west);

\node[draw, fill=blue!20, anchor=west, align=left, text width=8.5cm, minimum height=1.8cm, rounded corners=3pt, drop shadow] (auditbox) at (4.25,3.0)
{{\footnotesize
\tikz\draw[fill=white] (0,0) circle (0.05cm); \hspace{0.5ex} Standardized black-box evaluations usable across vendors\\[1ex]
\tikz\draw[fill=white] (0,0) circle (0.05cm); \hspace{0.5ex} Cryptographic commitments for external verification\\[0.75ex]
\tikz\draw[fill=white] (0,0) circle (0.05cm); \hspace{0.5ex} Independent certifiers for audit and compliance review
}};
\draw[-{Stealth}] (2.5,3.0) -- (auditbox.west);

\node[draw, fill=blue!35, anchor=west, align=left, text width=8.5cm, minimum height=1.8cm, rounded corners=3pt, drop shadow] (policybox) at (4.25,5.0)
{{\footnotesize
\tikz\draw[fill=white] (0,0) circle (0.05cm); \hspace{0.5ex} Required disclosures in system documentation\\[1ex]
\tikz\draw[fill=white] (0,0) circle (0.05cm); \hspace{0.5ex} Certification for deployment in high-governance domains\\[0.75ex]
\tikz\draw[fill=white] (0,0) circle (0.05cm); \hspace{0.5ex} Enforced penalties for unverifiable or non-compliant claims
}};
\draw[-{Stealth}] (1.5,5.0) -- (policybox.west);

\end{tikzpicture}
}
\caption{\textbf{A three-layer framework for enforceable watermarking.} Each layer represents a distinct governance function: technical guarantees, independent auditability, and regulatory enforcement. Arrows to the right point to concrete mechanisms that instantiate the requirements at each level, linking system design to policy accountability. Arrows to the left show the feedback between layers, with evidence flowing upward and revision flowing back down.}
\label{fig:watermarking-stack}
\end{figure*}

To bridge the gap between policy ambition and technical feasibility, we propose a three-layer framework: (I) technical requirements, (II) audit infrastructure, and (III) policy enforcement. Each layer targets a key weakness in current watermarking practices and together, they establish the groundwork for end-to-end accountability. Figure~\ref{fig:watermarking-stack} summarizes the framework.

The layers are functional roles rather than institutional silos or sequential stages. Existing regulatory regimes already separate the same three functions while running feedback between them (Table~\ref{tab:precedent}). The EU AI Act sets the requirements and the Commission asks CEN-CENELEC for standards. Systems that meet those standards are presumed compliant and what market surveillance finds goes back into the next revision~\cite{EUAIAct2024}. The one hard dependency is that a rule cannot reference a benchmark that does not exist, so the benchmarks have to be consolidated before regulators can set thresholds against them. Appendix~\ref{call-to-action} lays out that sequence and who is responsible for each step.

\begin{table}[h]
\centering
\small
\caption{Precedents for the three layers in existing regulatory regimes. In each, standards are adopted by reference and testing is independent of the regulated firm.}
\label{tab:precedent}
\begin{tabular}{@{}>{\raggedright\arraybackslash}p{3.3cm}>{\raggedright\arraybackslash}p{3.4cm}>{\raggedright\arraybackslash}p{3.2cm}>{\raggedright\arraybackslash}p{2.8cm}@{}}
\toprule
\rowcolor{lightgray}
\textbf{Layer} & \textbf{Cryptographic modules (NIST CMVP)}~\cite{FIPS1403} & \textbf{Medical devices (FDA)}~\cite{FDAASCA} & \textbf{AI systems (EU AI Act)}~\cite{EUAIAct2024} \\
\midrule
Layer 1:\newline Technical requirements & FIPS 140-3, adopting the ISO/IEC standard & FDA-recognized consensus standards & Harmonized standards \\
\midrule
Layer 2:\newline Audit infrastructure & Accredited private laboratories test; NIST validates & ASCA-accredited laboratories test; FDA accepts results & Notified bodies \\
\midrule
Layer 3:\newline Policy and enforcement & Procurement gating; public registry & Premarket review gates market entry & Market surveillance; penalties \\
\bottomrule
\end{tabular}
\end{table}

\subsection{Technical Requirements}
Effective governance depends on watermarking systems that are technically sound and externally testable. At a minimum, watermarking methods should demonstrate robustness to both benign transformations and adversarial modifications designed to evade detection. Systems must report standardized performance metrics, including detection accuracy, false positive rates, and robustness under defined perturbation sets. Crucially, these guarantees should not require full transparency of proprietary model internals but must expose interfaces and artifacts, such as detectors, keys, or hashes, that enable reproducible evaluation by third parties. This layer establishes the minimum technical properties needed to claim that watermarking is doing more than signaling intent.

Watermarking spans text, image, and audio~\cite{zhao2024sok}, and each modality exhibits specific transformations that risk watermark integrity. Text watermarks face paraphrasing, spelling alterations, lexical substitutions, summarization, and machine translation~\cite{kirchenbauer2023watermark}; image watermarks must withstand lossy compression (e.g., JPEG), geometric manipulations such as cropping, rotation, and scaling, color shifts, and style transfer~\cite{zhao2024invisible}; and audio watermarks contend with background noise injection, pitch shifts, re-encoding, and time-stretching~\cite{chen2023wavmark}. A watermarking scheme must be evaluated against these transformations not only for robustness, but also for recoverability without false attribution.

Benchmarking efforts such as MarkMyWords~\cite{piet2025markmywords} (text), WAVES~\cite{an2024waves} (images), and AudioMarkBench~\cite{liu2024audiomarkbench} (audio) provide standardized robustness evaluations across modalities, while a growing body of work examines attack vectors including removal, forgery, and spoofing~\cite{liang2025a, jovanovic2024watermark}. While these contributions offer valuable technical foundations, they remain research artifacts rather than governance infrastructure. Bridging this gap requires not new benchmarks per se, but the institutionalization of benchmarking within enforceable governance frameworks.

We suggest the creation of a shared benchmarking library for generative watermarking evaluation, accessible to industry, regulators, and auditors. Such a library would consolidate existing transformation pipelines standardized across modalities into centralized infrastructure with automated reporting of scores. Because watermarking cannot tolerate mislabeling human content as AI-generated, metrics must emphasize TPR at low false positive rates (e.g., 1\% or 0.1\% FPR). Where applicable, modality-specific metrics should also be reported.

Systems must also document tradeoffs, such as reductions in model output quality, inference latency increases, or decreased robustness to noise, as well as the specific watermark strength and threshold configurations intended for deployment. This transparency allows users and auditors to verify that certified parameters match production settings. To stay ahead of emerging threats, the library should also support contributions from red-teaming initiatives and adversarial research challenges, ensuring that benchmark suites evolve alongside attack capabilities~\cite{ahmad2025openai, anthropic2025frontier}.

While Executive Order 14110~\cite{ExecOrder14110} identified NIST as a natural home for watermarking benchmarks, the pace of AI deployment exceeds NIST's update cycles. We recommend NIST serve as registry and certifier while a community-driven consortium handles operational development, combining institutional legitimacy with rapid iteration. This follows existing federal practice where agencies are directed to use voluntary consensus standards in place of government-unique ones~\cite{OMBA119}, and NIST's own FIPS 140-3 adopts the ISO/IEC standard by reference~\cite{FIPS1403}.

To institutionalize this layer, any system claiming compliance must submit its watermarking pipeline to this library's API for evaluation. Recognizing potential concerns over intellectual property or security, alternative compliance paths such as exposing a detection API or providing zero-knowledge proofs could be considered, provided they allow equivalent third-party evaluation. Outputs will be scored on standardized perturbation suites, with results logged in a public registry. This shifts evaluation from self-reported metrics to verifiable compliance with shared robustness expectations.
These evaluations also form the empirical basis for audit infrastructure in Layer 2, enabling certifiers to assess compliance without needing privileged access.

\subsection{Audit Infrastructure}\label{audit-infra}
Robust watermarking requires more than technical performance as it must be verifiable through independent and reproducible evaluation. This layer establishes the infrastructure for third-party audits, including standardized black-box testing protocols, shared testbeds, and recognized certifying bodies. Audits should reflect real-world use conditions rather than idealized lab settings and must work across different vendors without requiring access to proprietary internals. Systems should expose externally testable behaviors or cryptographic commitments that allow verification without reverse engineering. By enabling interoperable, independent testing, this layer transforms watermarking from internal assurance into a publicly accountable mechanism.

Full public access to a detection system would enable adversaries to remove or forge watermarks, so key detection parameters must remain concealed. Consequently, audit systems must operate in a black-box setting where evaluators do not access model internals, but instead submit test content and observe binary or probabilistic detection outcomes. This approach introduces challenges since black-box detectors are vulnerable to repeated querying attacks, where adversaries iteratively probe a system to learn its boundaries~\cite{sadasivan2023can}. To mitigate this, detection interfaces must implement access controls, either via restrained queries or containerized deployments~\cite{tramer2016stealing}. These controls define two access tiers: public detection interfaces are rate-limited, whereas accredited auditors receive bounded query budgets in on-premise environments where the detector never leaves vendor control. Because detection query logs are themselves a surveillance surface, both tiers must be paired with purpose limitations and log-retention rules as certification criteria.

Watermarking deployments should also support cryptographic commitments that allow external verification without revealing sensitive internals. This can involve traditional hash-based attestations (e.g., SHA-256 commitments to watermark parameters)~\cite{pub2012secure} or emerging cryptographic techniques such as zero-knowledge proofs~\cite{sasson2014zerocash}. For example, a system might prove to an auditor that it conforms to a certified detection threshold, without disclosing the underlying keys or watermark design and without the auditor or the vendor inspecting user content, which also reduces the query budget an auditor needs.

A central question concerns auditor identity and governance. Potential certifiers include NIST-accredited laboratories, third-party security firms, or consortia established for compliance auditing~\cite{ai2023artificial}. Certification levels may include baseline compliance for systems meeting minimum robustness, interoperability certification for systems supporting cross-vendor detection, or adversarial resilience certification for schemes robust under threat-model testing. Interoperability certification covers multi-key mechanisms across models and platforms, while baseline certification is limited to provider- or model-level keys, since the governance target is provenance of the model rather than identity of the user, as Article 50 of the EU AI Act requires~\cite{EUAIAct2024}. Per-user or per-session keys constitute fingerprinting~\cite{boneh1998collusion} and require a separate legal basis and privacy review~\cite{GDPR2016, McIntyre1995}.

Audit cadence is equally critical. Evaluations should occur: (i) pre-deployment, to certify governance readiness, (ii) periodically post-deployment, to catch degradation or drift, and (iii) post-incident, in response to adversarial exploits or system failures. A fixed test protocol is not enough on its own, for example, Volkswagen's diesel engines passed the standard laboratory cycle for years and were caught only when a third party ran independent on-road tests~\cite{ICCT2014}. Auditors must be able to reproduce robustness metrics defined in Layer 1 using only externally exposed interfaces. This includes performance on benchmark perturbation sets, detection behavior under adversarial conditions, and public system documentation. Certified systems should be logged in a public registry indicating audit status and any revocations or compliance failures. A prototype evaluation scorecard illustrating these technical benchmarks and audit criteria is shown in Table~\ref{table1}.

\begin{sidewaystable*}[htbp]
\centering
\caption{Evaluation Scorecard for Layer 1 Watermarking Requirements}
\label{table1}
\vspace{1em}
\renewcommand{\arraystretch}{1.3}
\begin{tabular}{>{\bfseries}p{3.5cm} p{5.5cm} C{2.5cm} p{8.5cm}}
\toprule
\textbf{Category} & \textbf{Metric} & \textbf{Score Range} & \textbf{Comments / Benchmarks} \\
\midrule

\multirow{4}{=}{Robustness (Text)~\cite{math13091420}} 
  & Paraphrasing & 0--5 & 5 = survives 90\%+ of semantic paraphrasing tools \\
  & Token substitution & 0--5 & 5 = $<$5\% detection drop with synonym swaps \\
  & Translation & 0--5 & 5 = survives round-trip machine translation in 3+ languages \\
  & \dots & \dots & \dots \\

\midrule

\multirow{4}{=}{Robustness (Image)~\cite{an2024waves}} 
  & JPEG compression & 0--5 & 5 = survives down to 20\% quality with $\geq$95\% detectability \\
  & Cropping \& rotation & 0--5 & 5 = survives $\pm$10° rotation and 15\% central crop \\
  & Color/style shift & 0--5 & 5 = persists across style transfer and color remapping \\
  & \dots & \dots & \dots \\

\midrule

\multirow{4}{=}{Robustness (Audio)~\cite{wen2025sok}} 
  & Noise injection & 0--5 & 5 = survives pink/white noise @ 20 dB SNR \\
  & Time-stretching & 0--5 & 5 = survives $\pm$5\% tempo shift with detectable signal \\
  & \dots & \dots & \dots \\

\midrule

\multirow{4}{=}{Detection Quality} 
  & False Positive Rate & 0--5 & 5 = $<$0.1\% FPR on curated human-generated corpora \\
  & TPR @ Low FPR & 0--5 & 5 = $\geq$95\% TPR at 1\% FPR; $\geq$90\% TPR at 0.1\% FPR \\
  & False Negative Rate & 0--5 & 5 = $<$1\% FN under defined threat model \\
  & \dots & \dots & \dots \\

\midrule

\multirow{3}{=}{Auditability} 
  & Third-party reproducibility & 0--5 & 5 = Public test suite, verified by 2+ independent labs \\
  & Detection API/tool availability & 0--5 & 5 = Public API or SDK with access controls \\
  & \dots & \dots & \dots \\

\midrule

\multirow{3}{=}{Privacy} 
  & Key granularity & Yes/No & Must be provider- or model-level \\
  & Purpose limitation & 0--5 & 5 = Detection access restricted to certified purposes \\
  & Log retention & 0--5 & 5 = Detection query logs retained for a bounded period \\
  & \dots & \dots & \dots \\

\midrule

Modality Coverage & Text, Image, Audio support & 0--5 & 1 point per supported modality \\

Security Disclosure & Known failure modes documented & 0--5 & 5 = Threat model or model card made public \\

Policy Alignment & Standards-based declaration & Yes/No & Must align with NIST or equivalent standards framework \\
& \dots & \dots & \dots \\

\midrule

\dots & \dots & \dots & \dots \\

\bottomrule
\end{tabular}
\end{sidewaystable*}

\subsection{Policy and Enforcement}
The final layer ensures that technical and audit standards translate into real-world accountability. Without legal mandates and institutional enforcement, even robust and verifiable watermarking systems may go unused or implemented inconsistently. These mechanisms should be tied to existing governance processes, such as model registration or domain-specific safety requirements. Public disclosures, for instance, in model cards, help standardize expectations and support auditability. Certification should be required for deployment in high-risk contexts, and penalties should apply for unverifiable or misleading claims. This layer connects the technical and audit layers to institutional incentives, closing the loop from design to deployment.

Certification based on audit results should be required for deployment in high-governance contexts (e.g., elections, education). The robustness requirement itself should take the form of a testable condition such as a minimum true positive rate at a fixed low false positive rate on a named benchmark suite, with results verified by an accredited auditor. The metric and bar are shared across modalities for comparability, the perturbation suite is specific to each, and strictness scales with deployment risk (Appendix~\ref{appendix:thresholds}). Obligations attach to deployment context rather than developer size: every provider submits to the Layer 1 benchmark library and discloses results in its model card (Tier 0), whereas accredited audit (Tier 1) is paid for by the deployer that places a system into a high-governance context, on the vendor-pays accredited-laboratory model of NIST's CMVP~\cite{FIPS1403}, not by the upstream developer releasing open weights. This mirrors the EU AI Act, whose open-source exemption (Article 2(12)) does not extend to the Article 50 marking obligations or to high-risk placement~\cite{EUAIAct2024}; proportionality measures for small and open-source developers are given in Appendix~\ref{stakeholder-priorities}.

In tandem, mandated disclosures in system documentation and model cards must include audit status, robustness guarantees, known failure modes~\cite{mitchell2019model}, and the key granularity, log retention, and detection-access policy scored in Table~\ref{table1}. Non-compliant systems failing to meet audit standards or refusing audit participation should face graduated enforcement actions. These may include fines, removal of deployment licenses for regulated sectors, or public listing in non-compliance registries. Enforcement authority must be clearly defined and it does not need to sit with the body that keeps the standards. In the U.S., where NIST cannot regulate, NIST would hold the registry and certify, with enforcement running through procurement rules, as it already does for cryptographic modules~\cite{FIPS1403}. Designated regulators under the EU AI Act~\cite{EUAIAct2024} and China's Cyberspace Administration~\cite{CAC} can enforce within their jurisdictions. Since a model deployed in one jurisdiction produces outputs accessible in another, enforcement must also address international fragmentation. Under the Common Criteria Recognition Arrangement, a certificate from an accredited laboratory in one signatory is recognized across all members~\cite{CCRA2014}; watermarking certification can reuse this structure, with NIST, notified bodies under Article 43 of the EU AI Act~\cite{EUAIAct2024}, and the Cyberspace Administration as national certificate authorities recognizing one another's audit results.

Finally, just as audit protocols adapt to new adversarial techniques, enforcement mechanisms must include provisions for periodic policy review and revision. Feedback loops from auditors, researchers, and affected stakeholders can help keep regulatory frameworks responsive and legitimate. By linking technical performance and audit compliance to real-world consequences, Layer 3 closes the loop from design to deployment. It ensures that watermarking systems are not merely well-engineered, but meaningfully accountable in practice. Appendix~\ref{call-to-action} details dependencies and actions required to implement this framework.

\section{Alternative Views}\label{appendixB}
While we argue for enforceable watermarking as a critical step toward meaningful AI governance to transform it from a symbolic gesture into a mandatory mechanism, it is important to acknowledge opposing perspectives that highlight real implementation challenges.

Our proposed three-layer framework, though necessary for governance, poses substantial practical hurdles. The technical complexity alone requires the development of robust watermarking schemes across modalities, supported by standardized benchmarks and test protocols, and robustness against an unbounded adversary is provably unattainable~\cite{zhang2024watermarks} (Appendix~\ref{appendix:threat}). Beyond technical design, the framework calls for the creation of new institutions to oversee audit infrastructure and certification processes. These demands would foreseeably slow down the current rapid pace of generative AI innovation, particularly for smaller firms and open-source communities~\cite{kapoor2024societal, NTIA2024}, even with the tiering of Layer 3. The requirements are also dual-use: the durability and reliable detection Layer 1 demands are what a tracing infrastructure would want, and fingerprinting embeds per-recipient marks precisely to identify individuals~\cite{boneh1998collusion}. C2PA's own harm assessment lists deanonymization and tracking of content creators among the risks of provenance infrastructure~\cite{C2PAHarms}, and OpenAI reportedly withheld its text watermark partly over expected user backlash~\cite{wsj2024watermark}, the incentive described in Section~\ref{incentives}; this is why Layer 2 confines baseline certification to model-level keys.

Others would confine watermarking to a narrow supporting role, and we agree that it cannot carry provenance alone. NIST AI 100-4 recommends layering content credentials, watermarking, and post-hoc classifiers because each covers another's failure mode~\cite{NISTAI1004}; credentials are stripped by re-encoding and upload~\cite{C2PAHarms}, classifiers proved unreliable~\cite{openai_ai_text_classifier_2023}, and watermarks fail for non-cooperating generators, where platform labels and secure logs carry the load. The three layers apply to each family alike.

In addition to technical challenges, the fragmented landscape of both AI development and global regulation limits the interoperability and enforceability of watermarking systems. Proprietary models often adopt incompatible watermarking implementations, while open-source forks can bypass governance requirements entirely. On the regulatory side, jurisdictions such as the United States, European Union, and People's Republic of China are developing divergent standards, ranging from voluntary commitments to strict mandates, resulting in conflicting requirements. A model compliant in one region may not meet the standards of another, undermining global accountability efforts.

Amid this fragmentation, some would argue that voluntary compliance offers flexible and innovation-friendly alternatives to rigid enforcement~\cite{marchant2019soft, hagemann2018soft}. As noted in Section~\ref{ass3}, shifts in United States policy have demonstrated this volatility while one administration secured commitments from major AI firms, a subsequent administration rescinded those efforts. For some, this approach appears more politically viable than establishing binding standards. However, we contend that voluntary commitments alone are insufficient. In the absence of enforceable requirements and independent audits, companies are likely to implement minimal or symbolic watermarking measures that fall short of supporting meaningful oversight. The strongest alternative view is that overreliance on watermarking creates false confidence and distracts from literacy and governance, since flagging only some misleading content increases belief in the rest~\cite{pennycook2020implied}. We read this as an argument for enforceable standards rather than against watermarking, because false confidence is the product of such symbolic, unaudited watermarking, which Layer 3's mandated disclosure of failure modes is meant to prevent.

\section{Conclusion}
Watermarking is rapidly becoming a pillar of AI governance, but without enforceable standards and verifiable implementation, it risks becoming a symbol of oversight rather than a mechanism of accountability. We argue that robustness, verifiability, and auditability must be built into watermarking from the ground up and not added as compliance afterthoughts. Realizing this vision requires a structural shift for clear technical baselines, independent audit infrastructure, and regulatory enforcement that binds them together. Achieving durable accountability will depend on sustained collaboration between policymakers, industry actors, and the technical research community.

\section*{Acknowledgments}
This work was supported by the National Science Foundation (NSF) under grant numbers OAC-2112606, 2427505, and 2141622, by the National Library of Medicine of the National Institutes of Health (NIH) under award number R01LM014520, and by the U.S. Department of Agriculture (USDA) National Institute of Food and Agriculture under award number 2026-67023-46766.

\bibliographystyle{abbrv}
\bibliography{references}

\appendix

\section{Call to Action}
\label{call-to-action}
Section~\ref{three-layer-framework} specifies \textit{what} must change; this appendix addresses \textit{who} should act, \textit{when}, and in \textit{what sequence}. We distinguish near-term steps from medium- to long-term objectives and highlight dependencies across stakeholders.

\subsection{Sequencing and Dependencies}
Progress requires coordinated movement, but not all actors can move simultaneously. Standards bodies must move first, convening working groups and designating provisional benchmarks from existing academic infrastructure. As there exist current benchmarks in the literature, this foundational work is achievable within the first year. This enables policymakers to reference shared benchmarks in regulatory guidance rather than relying on undefined terms. Drawing from different benchmarking efforts, regulators can specify quantitative thresholds, for example, requiring at least 90\% true positive rates at 1\% false positive rates post-transformation, and industry actors can begin voluntary submission to evaluation suites. Enforcement mechanisms for high-risk domains and cross-jurisdictional mutual recognition agreements represent medium-term objectives, likely requiring two to four years of coordinated effort modeled on existing frameworks in product safety and financial regulation.

This sequencing reflects a practical constraint: enforceable mandates require shared benchmarks, and shared benchmarks require initial research and standards coordination.

\subsection{Stakeholder-Specific Priorities}\label{stakeholder-priorities}
Each stakeholder occupies a distinct position in the chain described in Section~\ref{three-layer-framework}, with corresponding strategic priorities.

\textbf{Standards Bodies.} 
Candidate benchmarks already exist, such as MarkMyWords~\cite{piet2025markmywords} for text, WAVES~\cite{an2024waves} for images, and AudioMarkBench~\cite{liu2024audiomarkbench} for audio. The near-term task is not to create new benchmarks but to consolidate these into a unified evaluation suite with standardized APIs and reporting formats. Organizations such as ISO/IEC JTC 1/SC 42 (the international standards committee in the area of AI), NIST, or domain-specific bodies like IEEE could serve as institutional homes. A lighter-weight alternative would be a multi-stakeholder consortium, modeled on MLCommons for ML benchmarking, that maintains the library while formal standards bodies provide certification authority.

\textbf{Policymakers.}
For example, the EU AI Act's requirement that watermarks be ``sufficiently robust...as far as this is technically feasible'' could be operationalized as follows: \textit{Watermarking systems deployed in high-risk contexts must demonstrate $\geq$90\% true positive rate at $\leq$1\% false positive rate on the [named benchmark suite, version X] perturbation tests, with results verified by an accredited auditor.} Such specificity transforms compliance from a matter of interpretation into a testable condition. Section~\ref{appendix:thresholds} specifies the per-modality suites and risk tiers and shows the condition is executable. Policymakers should also define escalation pathways for systems that fail certification, ranging from mandatory disclosure to deployment restrictions in regulated sectors.

\textbf{Industry.} 
Concrete near-term actions include: (1) publishing robustness metrics against standardized perturbation suites in model cards or system documentation; (2) exposing detection endpoints, whether public APIs with rate limiting or on-premise audit tools, that allow third-party verification without revealing proprietary keys; and (3) participating in pre-competitive consortia to develop shared detection infrastructure, reducing duplication and enabling interoperability. Firms that move early can also propose benchmark extensions reflecting real-world deployment conditions, ensuring that eventual standards reflect operational realities rather than purely academic threat models.

\textbf{Small and Open-Source Developers.}
Tier 0 is designed to cost little: the benchmark library is funded as a public good, so the marginal cost of submission approaches that of an upload, and marking itself is now cheap for open models since SynthID-Text ships in Hugging Face Transformers~\cite{synthid2024hf}. Openness is also an auditability asset, as the evaluation in Table~\ref{tab:watermark-eval} was possible only because Gemma and SynthID-Text are open, consistent with marginal-risk analyses of open foundation models~\cite{kapoor2024societal, NTIA2024}. For Tier 1, proportionality should follow existing practice: fee reductions and sandbox priority for SMEs under Article 62 of the EU AI Act~\cite{EUAIAct2024}, the FDA's small-business user-fee program~\cite{FDASmallBusiness}, and graduated assurance levels modeled on the security levels of FIPS 140-3~\cite{FIPS1403}.

\textbf{Researchers.} 
Two contributions fall squarely within the academic domain: (1) \textit{living benchmarks} that evolve with attack capabilities, preventing the obsolescence of static test suites; and (2) \textit{governance-aware design}, building watermarking schemes from inception to support auditability and interoperability rather than retrofitting these properties as compliance afterthoughts. Governance-aware design means, for instance, building watermarking schemes that output cryptographic commitments (e.g., hashes of detection parameters) at generation time, enabling later verification that a deployed system matches its certified configuration without exposing the parameters themselves.

\textbf{Funding.}
Realizing this infrastructure requires dedicated funding. Standards maintenance, benchmark updates, and independent audits are ongoing costs that neither academia nor individual firms are well-positioned to sustain alone. Public funding agencies (e.g., NSF, European Commission) and industry consortia should jointly resource this shared infrastructure, treating it as a public good analogous to measurement standards in other regulated domains.

No single stakeholder can close the governance gap alone. Standards enable audits, audits inform enforcement, and enforcement creates incentives for robust implementation. The framework succeeds only when each actor takes plausible steps within their domain while coordinating across dependencies.

\subsection{Threat Model}\label{appendix:threat}
The robustness claims in this paper are made against three tiers of transformation. T1 is routine, non-adversarial editing (e.g., compression, re-encoding, speed change, paraphrasing, summarization) by someone other than the generator, using a tool that could not have produced the content itself. T2 is resource-bounded adversarial modification aimed at the watermark (e.g., synonym substitution, targeted removal, and forgery or spoofing with a bounded query budget against the detector)~\cite{liang2025a, jovanovic2024watermark, sadasivan2023can}. T3 is the unbounded adversary, who can remove any watermark~\cite{zhang2024watermarks}, and is out of scope. Table~\ref{table1} and the certification condition of Appendix~\ref{appendix:thresholds} are scoped to T1 and T2, with baseline certification testing T1 and adversarial-resilience certification (Section~\ref{audit-infra}) testing T2. The evaluations of Section~\ref{govVtech} and Appendix~\ref{appendix:thresholds} are T1 only: the mandates quoted in Section~\ref{govVtech} are written against the content lifecycle rather than an attacker, so survival under routine edits is the compliance-relevant property.

A single-model evaluation cannot show what happens when content passes through a second generator, which is the usual case once text is reused, so we ran a two-hop experiment on text. Hop 1 is the calibrated SynthID-Text run of Appendix~\ref{appendix:thresholds}, restricted to a random half of its 500 C4 prompts, each with a watermarked and an unwatermarked \texttt{Gemma-2-9b-it} completion of at most 200 new tokens, paired on the same prompt; on this half the hop-1 detector reaches an AUC of .981 and 82.0\% TPR at 1\% FPR, against .981 and 86.6\% on all 500. Hop 2 passes each untransformed hop-1 completion through \texttt{Qwen2.5-7B-Instruct}~\cite{qwen2025qwen25} with the instruction ``Paraphrase the passage below. Keep the original order of the ideas, preserve all of the information, and keep the paraphrase about the same length as the original. Write only the paraphrase.'' followed by the passage, under multinomial sampling (temperature 1.0, top-$p$ 1.0, top-$k$ 0, 32 to 320 new tokens, seed 42), once with KGW~\cite{kirchenbauer2023watermark} on and once with it off, the two runs of each source sharing a seed. This yields four buckets of 250: $\mathrm{A}'$ (watermarked $\rightarrow$ KGW on), $\mathrm{A}''$ (watermarked $\rightarrow$ KGW off), $\mathrm{B}'$ (unwatermarked $\rightarrow$ KGW on), and $\mathrm{B}''$ (unwatermarked $\rightarrow$ KGW off). No output reached the length cap or floor; the pooled median is 164 Qwen tokens.

KGW uses the MarkLLM defaults~\cite{pan2024markllm}: $\gamma = 0.5$, $\delta = 2.0$, a one-token left-hash context, and a $z$-score over all tokens after the first. SynthID is scored by the mean $g$-value detector with $n$-gram length 5, 30 keys, two leaves, and non-distortionary masking, using the Gemma tokenizer. Both detectors run on all four buckets, and each threshold is the 99th percentile of that detector's own 500 matched negatives, so every rate in Table~\ref{tab:hop} is an operating point at approximately 1\% FPR: SynthID is calibrated on $\mathrm{B}' \cup \mathrm{B}''$ (threshold .520, coinciding with MarkLLM's default of .52) and KGW on $\mathrm{A}'' \cup \mathrm{B}''$ ($z = 1.91$, far below the fixed 4.0 of the original scheme).

\begin{table}[h]
\centering
\caption{Two-hop detection at approximately 1\% FPR. Each bucket holds 250 \texttt{Qwen2.5-7B-Instruct} paraphrases of hop-1 \texttt{Gemma-2-9b-it} completions. Thresholds are the 99th percentile of each detector's matched negatives: SynthID on $\mathrm{B}' \cup \mathrm{B}''$ (.520), KGW on $\mathrm{A}'' \cup \mathrm{B}''$ ($z = 1.91$).}
\label{tab:hop}
\begin{tabular}{@{}lllcc@{}}
\toprule
\rowcolor{lightgray}
\textbf{Bucket} & \textbf{Hop 1} & \textbf{Hop 2} & SynthID & KGW \\
\midrule
$\mathrm{A}'$  & SynthID       & KGW on  & 4.8\% & 43.2\% \\
$\mathrm{A}''$ & SynthID       & KGW off & 6.8\% & 0.8\% \\
$\mathrm{B}'$  & unwatermarked & KGW on  & 0.8\% & 39.2\% \\
$\mathrm{B}''$ & unwatermarked & KGW off & 1.2\% & 1.2\% \\
\bottomrule
\end{tabular}
\end{table}

The paraphrase reduces the first mark to near the floor whether or not the second model marks its output (4.8\% and 6.8\% in $\mathrm{A}'$ and $\mathrm{A}''$), consistent with the impossibility result for strong watermarking~\cite{zhang2024watermarks}. The second mark is recoverable only where the second model applied it (43.2\% and 39.2\% in $\mathrm{A}'$ and $\mathrm{B}'$, against 0.8\% and 1.2\% where it did not). The instruction preserves order, information, and length, so the task is low-entropy and the second mark is weak at this length and would not clear the 90\% bar of Appendix~\ref{appendix:thresholds}; its presence, not its strength, is the finding. Provenance is re-established at each hop only when every model in the chain marks its output, and one non-marking hop erases it. That is the case for enforcement that reaches every deployer in a pipeline (Layer 3), not only the first generator.

\subsection{Certification Thresholds}\label{appendix:thresholds}
Section~\ref{three-layer-framework} requires that the robustness condition be testable. We propose one metric, true positive rate at a fixed low false positive rate (Layer 1), evaluated on a perturbation suite specific to each modality: MarkMyWords for text, scored on outputs of at least 180 tokens~\cite{piet2025markmywords}, WAVES for images~\cite{an2024waves}, AudioMarkBench for audio~\cite{liu2024audiomarkbench}, and, provisionally, the VideoSeal evaluation protocol for video~\cite{fernandez2024videoseal} until a community benchmark exists. The bar is shared across modalities and scales with deployment risk, following the graduated assurance levels of FIPS 140-3~\cite{FIPS1403} and the risk tiers of the EU AI Act~\cite{EUAIAct2024}. A standard high-governance tier requires at least 90\% TPR at no more than 1\% FPR after transformation, and a critical tier requires at least 95\% TPR at no more than 0.1\% FPR. The metric and bar are therefore shared for comparability, tests differ by modality, and strictness follows the deployment context. For text, detectability also depends on output entropy, which can differ across the groups a system serves~\cite{raz2024authorship}, so TPR and FPR should be reported per domain of the text suite rather than pooled.

The condition is executable with the schemes evaluated in Section~\ref{govVtech}. For each scheme we set the threshold at the 99th percentile of 500 clean unwatermarked scores, which fixes the operating point at approximately 1\% FPR, and apply that threshold unchanged to 500 watermarked outputs before and after each transformation and to 500 unwatermarked outputs after the same transformation, counting scores above it as detected. These runs use the same models as Table~\ref{tab:watermark-eval}, with prompts drawn from C4, AudioCaps, and PartiPrompts and with watermarked and unwatermarked outputs paired on the same prompt. At this scale the text transformations were applied with open models, DIPPER for paraphrasing~\cite{krishna2023paraphrasing} and Pegasus-XSum for summarization~\cite{zhang2020pegasus}, rather than GPT-4o, since a generator-class editor conflates removing a watermark with regenerating the content, whereas these tools can only rewrite what they are given; the remaining transformations matched Table~\ref{tab:watermark-eval} except that image and video noise was applied at 20\,dB SNR. Table~\ref{tab:calibrated} reports the calibrated threshold, the clean and attacked true positive rates, and the attacked false positive rate.

Table~\ref{tab:watermark-eval} used each scheme's claimed threshold, whereas Table~\ref{tab:calibrated} uses the calibrated one, so the AudioSeal noise results that fail at 0.5 largely pass at 0.127, and the PixelSeal and VideoSeal calibrated points land near the binomial value of 0.578. Image and video schemes clear the 90\% bar under benign edits, while the deployed text and audio schemes fall well below it. This is the certification gap made concrete, as the bar must be a stated TPR at a stated FPR rather than a vendor assertion. The 1\% tier is characterizable from 500 negatives; the 0.1\% tier requires roughly 3,000 negatives by the rule of three and is a proposed target rather than a measured rate.

\begin{table}[h]
\centering
\caption{Calibrated thresholds (99th percentile of 500 clean unwatermarked scores, approximately 1\% FPR) and resulting rates. Thresholds are on each scheme's native score and are not comparable across rows; the summarization row uses 499 watermarked and 494 unwatermarked outputs after dropping seven summaries the detector could not score.}
\label{tab:calibrated}
\begin{tabular}{@{}llcccc@{}}
\toprule
\rowcolor{lightgray}
\textbf{Scheme} & \textbf{Transformation} & Threshold & Clean TPR & Attacked TPR & Attacked FPR \\
\midrule
SynthID-Text & Paraphrase & .515 & 86.6\% & 6.4\% & 2.4\% \\
SynthID-Text & Summarize & .515 & 86.6\% & 32.9\% & 27.5\% \\
AudioSeal & Speed & .127 & 100\% & 2.8\% & 3.2\% \\
AudioSeal & Noise & .127 & 100\% & 93.4\% & 0.0\% \\
PixelSeal & JPEG & .574 & 100\% & 100\% & 0.2\% \\
PixelSeal & Noise & .574 & 100\% & 100\% & 1.2\% \\
VideoSeal & Speed & .566 & 100\% & 100\% & 1.0\% \\
VideoSeal & Noise & .566 & 100\% & 97.8\% & 1.4\% \\
\bottomrule
\end{tabular}
\end{table}

\subsection{Audit Access}\label{appendix:access}
Section~\ref{audit-infra} requires audits to run black-box. This is sufficient for every property Table~\ref{table1} scores, since detection accuracy, false positive rate, and robustness under a perturbation suite are input-output properties where the auditor submits content and observes the detection outcome. White-box access is necessary for two properties: conformance, meaning that the deployed strength, threshold, and key match the certified configuration, and key security, meaning that key material is generated and stored as documented. Cryptographic commitments close most of the conformance gap, as a hash of the certified parameters published at certification and re-attested at audit time lets an auditor check that the deployed configuration is the certified one without seeing it~\cite{pub2012secure}, and zero-knowledge proofs extend this to properties of the configuration~\cite{sasson2014zerocash}. What remains is handled as in the CMVP~\cite{FIPS1403}, where the vendor submits the implementation to an accredited laboratory under non-disclosure, the laboratory tests it against the standard, and only the result enters the public registry. The vendor's internals never reach the regulator or the public, which is what makes white-box testing acceptable to commercial providers and is already accepted practice for the cryptographic modules those providers deploy.

\section{Prompts and Example Outputs for Watermark Robustness Evaluation}\label{appendix:eval}
To supplement the evaluation results shown in Table~\ref{tab:watermark-eval}, we provide detailed methodology and example outputs for the four modalities evaluated: text (Section~\ref{appendix:text}), audio (Section~\ref{appendix:audio}), image (Section~\ref{appendix:image}), and video (Section~\ref{appendix:video}). Code is available at: \url{https://github.com/ANCP2021/Watermarking-Without-Standards}.

\subsection{Text: Google SynthID-Text Evaluation}\label{appendix:text}

We evaluate Google's SynthID-Text watermarking scheme using the open-source MarkLLM framework~\citep{pan2024markllm}. Although SynthID-Text is deployed in Google's proprietary Gemini models, direct access to detection scores is unavailable through Gemini's chat interface or API; detection results are returned only as binary classifications without confidence values. This limitation also applies to Google's SynthID implementations for images and audio, which operate as black boxes providing only binary ``likely AI-generated'' outputs. The lack of granular detection scores in these modalities precludes the quantitative robustness analysis we conduct here. In contrast, the open-sourced SynthID-Text implementation exposes detection scores, enabling systematic evaluation under transformation.

For our evaluation model, we select \texttt{Gemma-2-9b-it}~\citep{team2024gemma}. This choice reflects three considerations: Gemma 2 is Google's relatively recent open-source model family, providing alignment with SynthID-Text's intended deployment context; the 9B parameter size balances model capability with computational accessibility; and the instruction-tuned (it) variant reflects realistic deployment settings where users interact with chat-optimized models. Following the MarkLLM framework~\cite{pan2024markllm}, we use its default detection threshold of 0.52, classifying scores at or above this value as watermarked.

Table~\ref{tab:prompt-overview} lists the five prompts used for text generation, spanning factual, creative, and informal content types. To simulate realistic scenarios where users reprocess LLM outputs before redistribution, each generated output was transformed using GPT-4o~\citep{hurst2024gpt} with the paraphrasing and summarization instructions shown in the table. Table~\ref{tab:full-output-comparison} presents complete examples (original, paraphrased, and summarized outputs) for $\text{Prompt}_1$, illustrating how semantically equivalent reformulations can degrade watermark detection.

\begin{table}[h]
\centering
\caption{Evaluation prompts and transformation instructions used in robustness testing. Each output was paraphrased and summarized using GPT-4o with the listed instructions.}
\label{tab:prompt-overview}
\renewcommand{\arraystretch}{1.3}
\begin{tabular}{@{} l p{10.5cm} @{}}
\toprule
\rowcolor{lightgray}
\textbf{Prompt ID} & \textbf{Prompt Text} \\
\midrule
$\text{Prompt}_1$ & Summarize the causes and impacts of climate change. \\
$\text{Prompt}_2$ & Write a product review for a product you recently used. \\
$\text{Prompt}_3$ & Draft an email apologizing for a delayed response. \\
$\text{Prompt}_4$ & Explain quantum computing to a high school student. \\
$\text{Prompt}_5$ & Tell a short story about a robot learning emotions. \\
\midrule
\rowcolor{lightgray}
\textbf{Transformation} & \textbf{Instruction Text} \\
\midrule
Paraphrase & Paraphrase the following text while preserving its original meaning as closely as possible. \\
Summarize & Summarize the following text into 2-3 sentences, keeping the most important information. \\
\bottomrule
\end{tabular}
\end{table}

\subsection{Audio: Meta AudioSeal Evaluation}\label{appendix:audio}
We evaluate Meta's AudioSeal~\citep{roman2024proactive}, a post-hoc audio watermarking scheme designed for proactive detection of AI-generated speech and audio. Meta has deployed watermarking across its AI products for images, video, and audio content; however, while watermarks are reportedly applied to generated media, the specific schemes used for images and video are not publicly documented. AudioSeal is the only scheme Meta has explicitly identified as being applied to audio generation in their public-facing AI products~\citep{fernandez2025audioseal}.

For audio generation, we use \texttt{AudioGen}~\citep{kreuk2022audiogen}, an open-source text-to-audio model. This choice parallels our reasoning for selecting Gemma in the text evaluation: \texttt{AudioGen} was evaluated in the original AudioSeal work and represents an accessible proxy for proprietary audio generation systems where AudioSeal may be deployed. Each audio sample was generated as a 10-second clip and saved in WAV format, as specified by the AudioSeal implementation requirements. Table~\ref{tab:audio-prompt-overview} lists the five text descriptions used as prompts for \texttt{AudioGen}. Following the original AudioSeal implementation~\citep{roman2024proactive}, we use a detection threshold of 0.5, where scores at or above this value indicate watermark presence.

The selected transformations (1.25$\times$ playback speed and 10dB noise addition) simulate realistic scenarios where users apply simple modifications to audio content. Speed adjustment is common when sharing clips on social media or adjusting for time constraints, while background noise may be introduced through re-recording, compression artifacts, or environmental interference. Importantly, these transformations are deliberately non-adversarial; they represent routine user behavior rather than sophisticated attacks designed to evade detection. As shown in Table~\ref{tab:watermark-eval}, even these benign modifications substantially degrade watermark detectability, with speed modification proving particularly effective at evading detection across all samples.

\begin{table}[h]
\centering
\caption{Audio evaluation prompts (text descriptions for AudioGen) and transformation parameters. Each 10-second audio clip was saved as WAV and subjected to speed modification and noise addition.}
\label{tab:audio-prompt-overview}
\renewcommand{\arraystretch}{1.3}
\begin{tabular}{@{} l p{10.5cm} @{}}
\toprule
\rowcolor{lightgray}
\textbf{Sample ID} & \textbf{Prompt Text} \\
\midrule
$\text{Sample}_1$ & Heavy rain and thunder during a severe storm with wind blowing \\
$\text{Sample}_2$ & Coffee machine brewing with steam hissing and liquid pouring into a mug \\
$\text{Sample}_3$ & Office sounds with keyboard typing, mouse clicking, and a phone ringing \\
$\text{Sample}_4$ & Electronic equipment humming with beeps and digital processing sounds \\
$\text{Sample}_5$ & Footsteps walking slowly on a wooden floor then a door creaking open \\
\midrule
\rowcolor{lightgray}
\textbf{Transformation} & \textbf{Parameters} \\
\midrule
Speed Modification & Playback speed increased to 1.25$\times$ original \\
Noise Addition & Additive white noise at 10\,dB SNR \\
\bottomrule
\end{tabular}
\end{table}

\subsection{Image: Meta PixelSeal Evaluation}\label{appendix:image}
We evaluate Meta's PixelSeal~\citep{soucek2025pixelseal}, a post-hoc image watermarking scheme that embeds a 256-bit payload. As noted in Section~\ref{appendix:audio}, Meta has not publicly documented the schemes applied to images and video in its products, and no deployed image or video scheme publicly exposes detection scores. PixelSeal and VideoSeal are Meta's open post-hoc schemes for these modalities and therefore serve as accessible proxies, paralleling our use of Gemma and AudioGen.

For image generation, we use \texttt{SDXL}~\citep{podell2024sdxl} (\texttt{stable-diffusion-xl-base-1.0}). Each image was generated at 1024$\times$1024 resolution with 30 denoising steps and a classifier-free guidance scale of 5.0, and PixelSeal was then applied post-hoc with a 256-bit payload. Table~\ref{tab:visual-prompt-overview} lists five prompts which describe scenes used for audio generation rewritten as visual descriptions.

Neither PixelSeal nor VideoSeal publishes a deployment threshold, and both report detection as bit accuracy, the fraction of the 256 payload bits recovered. We therefore set the threshold by a binomial criterion. Under the unwatermarked null each bit is recovered with probability 0.5, so the number of correctly recovered bits follows $\text{Binomial}(256, 0.5)$, and 148 is the smallest count with $p \leq .01$. Dividing by 256 gives a bit accuracy of 0.578, which we apply to both modalities and which corresponds to an operating point of roughly 1\% false positives.

The selected transformations (JPEG compression at quality 50 and additive white Gaussian noise at 10\,dB SNR) parallel the audio evaluation and simulate routine handling of images, as JPEG re-encoding is applied by nearly every platform on upload and noise approximates re-capture or low-light artifacts. As shown in Table~\ref{tab:watermark-eval}, PixelSeal remained above threshold on all samples under both transformations, with JPEG compression leaving bit accuracy nearly unchanged and noise reducing it to between 0.613 and 0.738.

\subsection{Video: Meta VideoSeal Evaluation}\label{appendix:video}
We evaluate Meta's VideoSeal~\citep{fernandez2024videoseal} (v1.0), a post-hoc video watermarking scheme that embeds a 256-bit payload across frames and is detected by the same bit-accuracy criterion as PixelSeal.

For video generation, we use \texttt{CogVideoX-2b}~\citep{yang2024cogvideox}, an open-source text-to-video model. Each clip was generated as 49 frames at 8 frames per second at the pipeline default resolution of 720$\times$480, with 50 denoising steps and a guidance scale of 6.0, using the same five prompts as the image evaluation (Table~\ref{tab:visual-prompt-overview}). We use the 0.578 bit-accuracy threshold derived in Section~\ref{appendix:image}.

The selected transformations (1.25$\times$ playback speed and white Gaussian noise at 10\,dB SNR) mirror the audio evaluation. Speed modification was implemented by dropping frames, from 49 to 39 frames at the same frame rate (approximately 4.9 seconds), as a user or platform would when shortening a clip. As shown in Table~\ref{tab:watermark-eval}, VideoSeal survived speed modification on all samples with bit accuracy at or near 1.00 and survived noise on four of five, with $\text{Sample}_2$ falling to 0.504, just below threshold.

\begin{table}[h]
\centering
\caption{Image and video prompts (shared by SDXL and CogVideoX-2b) and transformation parameters. The prompts describe the five audio scenes of Table~\ref{tab:audio-prompt-overview} rewritten as visual descriptions.}
\label{tab:visual-prompt-overview}
\renewcommand{\arraystretch}{1.3}
\begin{tabular}{@{} l p{8.7cm} @{}}
\toprule
\rowcolor{lightgray}
\textbf{Sample ID} & \textbf{Prompt Text} \\
\midrule
$\text{Sample}_1$ & Heavy rain and thunder during a severe storm, wind bending the trees \\
$\text{Sample}_2$ & A coffee machine brewing, steam hissing as liquid pours into a mug \\
$\text{Sample}_3$ & An office desk with hands typing on a keyboard, a mouse and a ringing phone \\
$\text{Sample}_4$ & Racks of electronic equipment in a server room with blinking indicator lights \\
$\text{Sample}_5$ & A person walking slowly across a wooden floor toward a creaking open door \\
\midrule
\rowcolor{lightgray}
\textbf{Transformation} & \textbf{Parameters} \\
\midrule
JPEG Compression (image) & Quality factor 50 \\
Speed Modification (video) & Playback speed increased to 1.25$\times$ original by frame dropping (49 to 39 frames at 8 fps) \\
Noise Addition (image and video) & Additive white Gaussian noise at 10\,dB SNR \\
\bottomrule
\end{tabular}
\end{table}

\begin{table}[h]
\centering
\caption{Full outputs for original, paraphrased, and summarized versions of $\text{Prompt}_1$. 
The original text includes markdown-style formatting (e.g., \#\#, \textbf{**}) (i.e., from ChatGPT's copy function).}

\label{tab:full-output-comparison}
\renewcommand{\arraystretch}{1.3}
\begin{adjustbox}{max width=\textwidth}
\begin{tabular}{@{} l p{12cm} @{}}
\toprule
\multicolumn{2}{l}{\textbf{$\text{Prompt}_1$: Summarize the causes and impacts of climate change.}} \\
\midrule
\rowcolor{lightgray}
Original & \small \#\# Climate Change: Causes \& Impacts
\newline

**Causes:**
\newline

* **Greenhouse Gas Emissions:** Primarily from burning fossil fuels (coal, oil, gas), deforestation, and industrial processes. These gases trap heat in the atmosphere, leading to a warming effect.

* **Human Activities:** Agriculture, transportation, and energy production significantly contribute to greenhouse gas emissions.

*   **Deforestation:** Trees absorb CO2, so cutting them down reduces the planet's ability to absorb these gases.
\newline

**Impacts:**
\newline

*   **Rising Global Temperatures:**  Leading to more frequent and intense heatwaves, droughts, and wildfires.

*  **Sea Level Rise:**  Melting glaciers and thermal expansion of water threaten coastal communities and ecosystems.

*    **Extreme Weather Events:**  Increased frequency and intensity of storms, floods, and droughts, causing widespread damage and displacement.

*     **Ocean Acidification:**  Absorption of CO2 by oceans makes them more acidic, harming marine life and ecosystems.\\
\rowcolor{lightblue}
Paraphrased & \small \#\# Climate Change: Causes and Effects
\newline 

**Causes:**
\newline

* **Greenhouse Gas Emissions:** Mostly result from burning fossil fuels such as coal, oil, and natural gas, as well as from industrial activities and deforestation. These gases trap heat in the Earth’s atmosphere, causing it to warm.

* **Human Activities:** Sectors like agriculture, transportation, and energy generation are major contributors to the release of greenhouse gases.

* **Deforestation:** Trees play a key role in absorbing carbon dioxide. When forests are cleared, this natural absorption capacity is diminished, increasing atmospheric CO2 levels.
\newline

**Effects:**
\newline

* **Increasing Global Temperatures:** This leads to more intense and frequent heatwaves, wildfires, and periods of drought.

* **Rising Sea Levels:** Caused by melting ice and the expansion of warmer ocean water, posing risks to coastal regions and habitats.

* **More Severe Weather Events:** Storms, floods, and droughts are becoming more common and intense, resulting in significant destruction and displacement of populations.

* **Ocean Acidification:** Oceans absorb large amounts of carbon dioxide, which lowers their pH and negatively affects marine organisms and ecosystems. \\
\rowcolor{lightyellow}
Summarized & \small Climate change is primarily driven by human activities such as burning fossil fuels, deforestation, and industrial processes, which release greenhouse gases that trap heat in the atmosphere. Its impacts include rising global temperatures, more extreme weather events, sea level rise, and ocean acidification, all of which pose serious threats to ecosystems and human communities. \\
\bottomrule
\end{tabular}
\end{adjustbox}
\end{table}

\end{document}